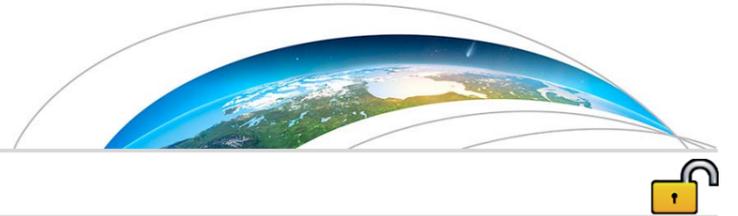

# Earth and Space Science

# A large-scale view of Space Technology 5 magnetometer response to solar wind drivers

D. J. Knipp[1,2], L. M. Kilcommons[1], J. Gjerloev[3,4], R. J. Redmon[5], J. Slavin[6], and G. Le[7]

[1]Aerospace Engineering Sciences, University of Colorado Boulder, Boulder, Colorado, USA, [2]High Altitude Observatory, NCAR, Boulder, Colorado, USA, [3]Applied Physics Laboratory, Johns Hopkins University, Laurel, Maryland, USA, [4]Birkeland Centre of Excellence, University of Bergen, Bergen, Norway, [5]National Geophysical Data Center, NOAA, Boulder, Colorado, USA, [6]Atmospheric, Oceanic and Space Sciences, University of Michigan, Ann Arbor, Michigan, USA, [7]NASA Goddard Space Flight Center, Greenbelt, Maryland, USA

**Abstract** In this data report we discuss reprocessing of the Space Technology 5 (ST5) magnetometer database for inclusion in NASA's Coordinated Data Analysis Web (CDAWeb) virtual observatory. The mission consisted of three spacecraft flying in elliptical orbits, from 27 March to 27 June 2006. Reprocessing includes (1) transforming the data into the Modified Apex Coordinate System for projection to a common reference altitude of 110 km, (2) correcting gain jumps, and (3) validating the results. We display the averaged magnetic perturbations as a keogram, which allows direct comparison of the full-mission data with the solar wind values and geomagnetic indices. With the data referenced to a common altitude, we find the following: (1) Magnetic perturbations that track the passage of corotating interaction regions and high-speed solar wind; (2) unexpectedly strong dayside perturbations during a solstice magnetospheric sawtooth oscillation interval characterized by a radial interplanetary magnetic field (IMF) component that may have enhanced the accompanying modest southward IMF; and (3) intervals of reduced magnetic perturbations or "calms," associated with periods of slow solar wind, interspersed among variable-length episodic enhancements. These calms are most evident when the IMF is northward or projects with a northward component onto the geomagnetic dipole. The reprocessed ST5 data are in very good agreement with magnetic perturbations from the Defense Meteorological Satellite Program (DMSP) spacecraft, which we also map to 110 km. We briefly discuss the methods used to remap the ST5 data and the means of validating the results against DMSP. Our methods form the basis for future intermission comparisons of space-based magnetometer data.

## 1. Background and Motivation

Space magnetometer measurements of field-aligned currents (FACs) in Earth's high-latitude regions are a vital tool for understanding the electrodynamics and heating mechanisms of the low Earth orbit environment. The New Millennium Program, Space Technology 5 (ST5) mission produced ~5000 polar passes through Earth's FAC system during 27 March to 27 June 2006. The magnetic data were previously available without gain-jump corrections and only in ascii-formatted SM coordinates. We created an enhanced space-based magnetometer data set from the mission that is newly archived in the NASA CDAWeb virtual observatory (VO) and described in this data report. The enhanced data have been jump corrected, referred to a common altitude, and placed in a common data format (NASA CDF) at http://cdaweb.sci.gsfc.nasa.gov/cdaweb/sp_phys/. Herein we describe how the amended data have been cross-checked against magnetic observations from the Defense Meteorological Satellite Program (DMSP) spacecraft. This mapping and validation exercise is a prelude to a much larger effort in reprocessing a decade's worth of magnetometer data from the DMSP for inclusion in a NASA VO. Thus, the validation methods we discuss here serve to address the ST5 data in particular and provide a path forward in improving other space-based magnetometer data sets. To provide an overview of the data effort, we show how the full-mission ST5 data can be visualized as a magnetic "keogram" and compared to the passage of high-speed solar wind streams and interspersed solar wind transients.

The ST5 mission consisted of three 25 kg spacecraft launched into 105.6° inclination, Sun-synchronous, 300 km × 4500 km, pearls-on-a-string orbits with periods of 136 min [*Slavin et al.*, 2008]. Each satellite hosted a boom-mounted triaxial fluxgate magnetometer. Flying in formation with separations of 50–5000 km, the spacecraft regularly crossed FAC structures during their 90 day mission. The interval was at the transition into the solar minimum between solar cycles 23 and 24. Thus, it provides a (mostly) geomagnetically quiet backdrop for the validation discussed herein. In the Northern Hemisphere the spacecraft sampled mostly sunward of the dawn-dusk meridian, while in the Southern Hemisphere the spacecraft sampled mostly tailward of the dawn-dusk meridian (see Figures 1a and 1b). The constellation operated in an experimental autonomous mode after 1 June 2006, and in a full "lights out" simulation mode from 11 to 18 June 2006, resulting in a lower rate of data capture (~60%) toward the end of the mission.

Although limited in duration, the high-quality data from the ST5 mission have yielded important insight into the dynamics of quiet-to-moderate activity-level FACs. *Slavin et al.* [2008] derived and intercompared, for the first time, field-aligned current densities using the standard single-spacecraft [e.g., *Zanetti et al.*, 1984] and two-point gradiometry methods. The FAC thickness, motion, and current intensity were found to be very stable for the small number of events considered and the current densities derived by the two techniques agreed to within ~ 10%. *Wang et al.* [2009] used the ST5 data to determine the speed of FAC motion and evaluated the effects of FAC sheet speed on FAC thickness and current density calculations. In case studies, *Le et al.* [2009] showed that mesoscale current structures, often found within large-scale field-aligned current sheets, have highly variable current density and/or polarity on time scales of ~ 10 min but appear to be relatively stable at the ~ 1 min time scale. *Cumnock et al.* [2011] studied transpolar arcs with multipoint ST5 measurements and found similar temporal variability. While investigating larger-scale variations, *Le et al.* [2010] determined that the imbalance of the Region 1 and Region 2 currents required Pedersen closure currents across the polar cap with magnitude of ~0.1 MA. *Le et al.* [2011] showed that the ST5 dawn-dusk orbits at appropriate altitudes and magnetic latitudes reveal azimuthal characteristics of field line resonances associated with Pc2–Pc3 waves. In a statistical study of the ST5 database, *Gjerloev et al.* [2011] reported that dayside and nightside magnetospheric reconfiguration times were ~1 min and ~3 min, respectively, based on FAC stability.

Most of the aforementioned studies were undertaken with the ST5 data provided in the solar magnetospheric (SM) coordinate system. Future researchers may find the data more amenable to analysis if the data are available at a common reference altitude and in other geomagnetic coordinates (see *Laundal and Gjerloev* [2014] for discussion of alternative magnetic coordinates). In this study we project the ST5 data in Modified Apex (Apex) coordinates [*Richmond*, 1995] to 110 km, an altitude near which ionospheric $E$ region conductivity tends to maximize. Doing so facilitates the interhemispheric data comparisons in sections 3 and 4.

In a keogram-view of the full-mission magnetic record we show the expected magnetic local time organization of field aligned currents (FACS) and FAC enhancements associated with solar wind disturbances, as well as long-lived magnetic calms associated with intervals of slow solar wind. We also note an unanticipated enhancement in the dayside high-latitude current systems during a solstice sawtooth oscillation event and subsequent high-speed solar wind driving. We suggest that ST5 measured effects of the radial interplanetary magnetic field (IMF)-magnetosphere interactions that are unique to the summer high-latitude dayside region. This data report is organized as follows: In section 2 we explain the data processing. In section 3 we provide the geophysical context for the ST5 mission. In section 4 we discuss the factors leading to diminutions and enhancements of FACs. We summarize our results in section 5.

## 2. Methods

### 2.1. Data Processing

Research-grade miniature fluxgate magnetometers onboard each ST5 spacecraft were deployed on a boom at ~1.1 m from the spacecraft center to provide magnetic isolation from the spacecraft bus. *Slavin et al.* [2008] found no evidence of stray magnetic field contamination from spacecraft operations and reported that instrument gains and offsets were stable, changing by less than 0.1% over the course of the mission. Each spacecraft processed its magnetometer data in real time using two dynamic ranges: ±64,000 nT, with 1.3 nT resolution (full-field) or ±16,000 nT, with 0.3 nT resolution (low-field). Magnetic vectors were generated at a 16 Hz rate (every 62.5 ms) [*Slavin et al.*, 2008]. Uncertainties in spacecraft absolute position varied during the mission but typically ranged from 1 to 5 km [*Purucker et al.*, 2008]. More details about the mission and the magnetometers can be found in *Carlisle et al.* [2006] and at https://directory.eoportal.org/web/eoportal/satellite-missions/s/st5.

Daily text files of ST5 magnetometer data contain average 1 s cadence, vector magnetic perturbation measurements along with the International Geophysical Reference Field (IGRF) 10 [*Maus and Macmillan*, 2005] vectors at the location of each datum. Both sets of vectors are provided in solar magnetospheric (SM) coordinates. At middle and low latitudes the original data set showed baseline jumps in all spacecraft and all three components on most orbits. For their comprehensive statistical study of the FAC variability, *Gjerloev et al.* [2011] performed a manual correction of middle- and high-latitude data aimed at correcting these jumps. They removed baseline shifts by assuming the values on each side of the jump to be identical. Strictly speaking the field could change over the period between the two data points, but this was assumed a small error compared to the jump itself. These corrected data are included in the VO files. We compared both the original ST5 data set with gain jumps, and the Gjerloev et al. corrected data set against Defense Meteorological Satellite Program (DMSP) spacecraft magnetic perturbations and found better comparisons with DMSP using the corrected data. Thus, for this study we use the Gjerloev et al. corrected version of the data.

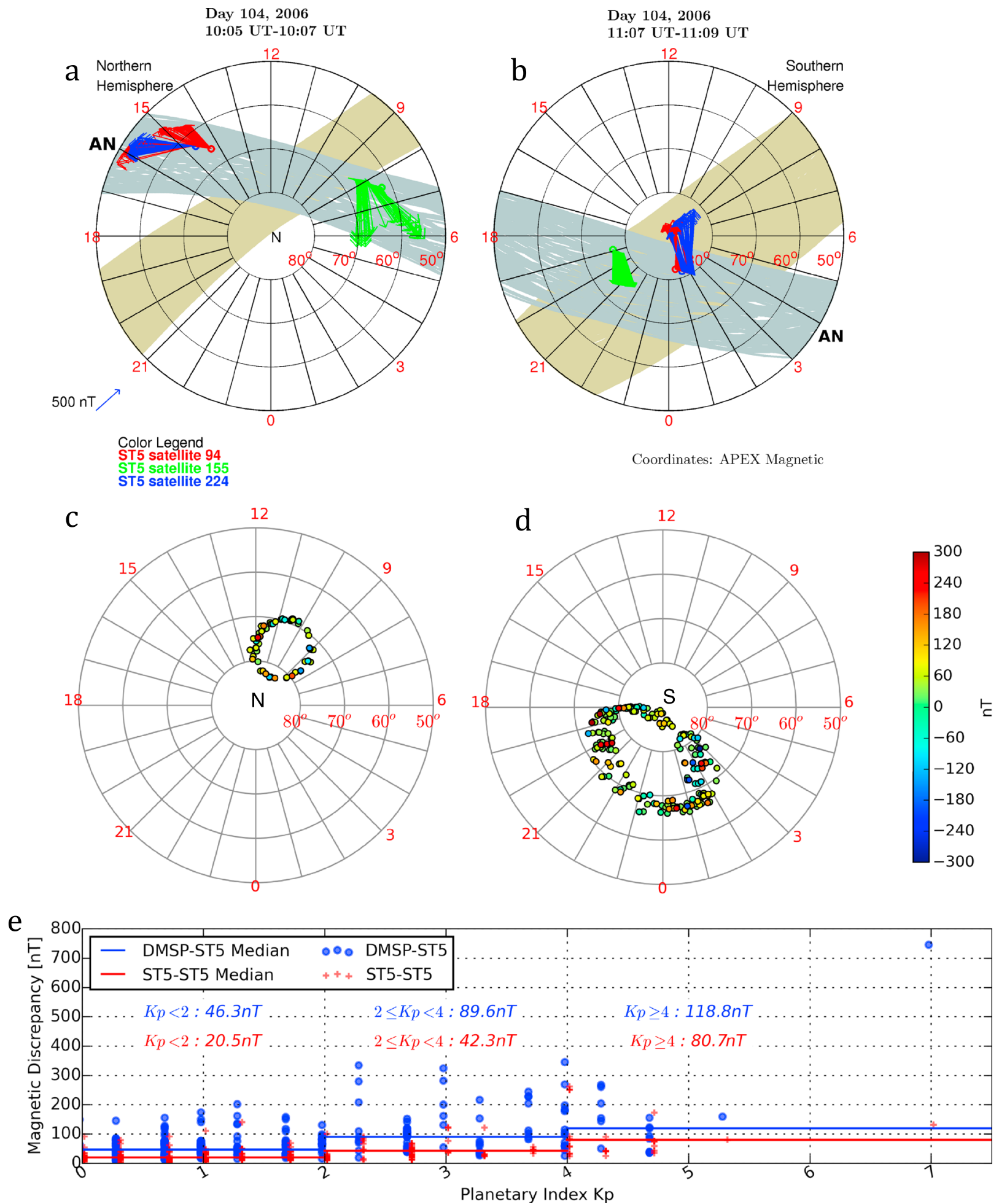

**Figure 1.** Coverage and comparison for the ST5 mission. Orbit coverage is in gray for ST5 and olive for DMSP F15 and F16. (a) Northern Hemisphere (NH) and (b) Southern Hemisphere (SH), and the local time of the ascending node is indicated by AN. Perturbation vectors from the individual ST5 spacecraft from a 2 min sequence on 14 April 2006 are shown in red ST5-94, green ST5-155, and blue ST5-224; (c and d) NH and SH ST5-DMSP magnetic discrepancies for magnetic conjunctions during 25 March to 26 June 2006. The conjunctions occurred within ±60 s and 3.0° in Modified Apex coordinates. The view is (Figure 1a) from above the NH and (Figure 1b) from inside of Earth looking toward SH. Positive values (warm colors) indicate discrepancies where DMSP > ST5. Negative values (cool colors) indicate discrepancies where ST5 > DMSP; (e) discrepancy magnitude versus Kp, DMSP-ST5 discrepancies are blue, ST5-ST5 discrepancies are red. Horizontal lines indicate median values in the Kp bins.

Table 1. Coordinates Systems for the ST5 Spacecraft Location and Magnetic Data

| Coordinate System | Spacecraft Location | Measured **B** Field | IGRF Model Field | **B** Field Perturbations |
|---|---|---|---|---|
| Solar magnetospheric (SM) | Yes | Yes | Yes | Yes |
| Geographic (GEO) | Yes | No | No | Yes |
| Modified Apex (Apex) | Yes | No | No | Yes |
| Altitude Adjusted Corrected Geomagnetic Coordinates | Yes | No | No | No |

After baseline correction and IGRF model-field removal, the high-latitude residual of the ST5 magnetic field intensity is approximately zero except for three situations: (1) when the spacecraft encounter FAC structures, (2) when the spacecraft encounter Pc waves, and (3) during some Northern Hemisphere perigee passes when perturbations associated with horizontal ionospheric currents were measured [*Slavin et al.*, 2008]. Depending on the local time of the orbit, during the perigee passes the ionospheric currents may enhance or reduce the magnetic field [*Zanetti et al.*, 1984].

The jump-corrected ST5 data were transformed into Earth Centered Inertial, Geocentric (GEO), and Modified Apex coordinates. (Table 1 provides a brief description of the contents of the CDF files.) The Apex transform allows the magnetic data to be scaled to different altitudes for comparison with data from other spacecraft constellations, referred to the same altitude. We performed both visual inspections and statistical analysis (discussed in section 2.2) of the data to ensure that the gain-jump removals and coordinate transformations were producing appropriate results. As in *Richmond* [1995] and *Knipp et al.* [2014] for the Apex transforms, we made the approximation that the magnetic perturbations orthogonal to the main geomagnetic field map along field lines in the same manner as electric fields. Mapping from higher satellite altitudes to 110 km in Modified Apex coordinates produces an increase in perturbation strength roughly proportional to $[(R_E + \text{Satellite Altitude})/(R_E + 110)]^{3/2}$. This increase accounts for the changes in FAC intensity caused by convergence of the magnetic field lines [e.g., *Mozer*, 1970; *Rich et al.*, 1981]. The reader is referred to equations (1)–(8) in *Knipp et al.* [2014] for the mathematical summary of Apex coordinates and to http://www.ngdc.noaa.gov/geomag/geom_util/apex.shtml for additional details. Full details can be found in *Richmond* [1995].

In this report we discuss two forms of magnetic data: (1) Perturbation vectors, which are the residual magnetic vector after main field and baseline removal, and (2) discrepancy vectors, which are the vector difference between two spacecraft measurements of (nearly) the same perturbation vector. Most of our graphics show the magnitude of the discrepancy vector. In an ideal world the perturbation vectors would be due to FACs and Pc waves and the discrepancy vector values would be zero. Figures 1a and 1b illustrate a few of the remapped ST5 horizontal perturbation vectors from an active interval on 14 April 2006 (Day 104). The spacecraft follow the same trajectory separated by 1 to 10 min depending on orbital location. The observations in Figure 1a are in the Northern Hemisphere near orbit perigee between 10:05 and 10:07 UT. In the snapshot spacecraft 155 is the lead spacecraft. It sampled the dawnside FACs (with maximum magnetic perturbations of ~950 nT) while the trailing satellites, 094 and 224 in sequence, sampled the afternoon FACs (with maximum perturbations of ~850 nT). Measurements from the subsequent Southern Hemisphere pass were made near apogee during the interval 11:07–11:09 UT (Figure 1b). The nearly uniform perturbations in the Southern Hemisphere are consistent with fields generated by FACs in the distant auroral zone. The combination of orbital eccentricity and mapping into Apex coordinates produce the hemispheric difference in spacecraft separation.

### 2.2. Interconstellation and Intraconstellation Comparisons

To verify that we were treating the ST5 data correctly, we compared the ST5 data with similarly processed data from DMSP spacecraft. Information on the DMSP magnetometer can be found in *Alken et al.* [2014]. The DMSP spacecraft tracks are shown in olive colors in Figures 1a and 1b. These data are also baseline corrected. Specifically, DMSP horizontal magnetic perturbations are determined by subtracting the appropriate epoch IGRF main field and a baseline value from each 1 s DMSP magnetic field datum. A baseline subtraction is needed to adjust for residual spacecraft magnetic signal, on-orbit sensor misalignment, and lack of on-orbit calibration. In accord with *Rich et al.* [2007] and *Knipp et al.* [2014], we performed a least squares polynomial fit to the data components observed while each satellite was at subauroral latitudes. We applied seventh- and fifth-order polynomials to the along-track and across-track component measurements, respectively, and calculated a root-mean-square (RMS) difference between the polynomial fits and original data. If the RMS error exceeded 15 nT (along-track) and 20 nT (across-track), the fitting boundary was reset ~10° equatorward of the auroral electron boundaries and the fitting was redone. If the RMS difference still exceeded the specified limits, the particular half-orbit of data was excluded from further processing.

We searched for subsets of the ST5 and DMSP data conjunctions within a window of 3.0° radius and ±60 s, using the DMSP orbit tracks as the basis. The window was chosen to be consistent the FAC stability limits reported by *Le et al.* [2009] and *Gjerloev et al.* [2011]. Although the orbit planes and parameters were quite different for DMSP (~14 nearly circular orbits/d) and ST5 (~10.5 elliptical orbits/d), we found 281 conjunctions in the orbit overlap regions (Figures 1c and 1d). From the point pairs within the conjunction window, the search algorithm located a closest approach pair. A subsequent processing step searched for conjunction pairs within a small zone (0.1° great-circle distance, typically 20–30 data points) around the point of closest approach and produced data for statistical comparisons of the magnetic discrepancies. The statistics determined include mean and median values and interquartile range of the discrepancies within the small zone. We also computed ST5-ST5 discrepancies within those same conjunction windows.

Most of the DMSP-ST5 conjunctions, 208/281, occur in the Southern Hemisphere where the constellation-orbit overlap and altitudinal separation is largest. Colored dots in Figures 1c and 1d indicate the magnitude of the median horizontal magnetic discrepancy in the small conjunction region of ST5 and DMSP measurements. For all 281 conjunctions the median of the discrepancy values is 57 nT. The median discrepancy value from the two closest ST5 spacecraft, using the same conjunction window, is ~ 30 nT. Since the closest ST5 spacecraft visit the same location with only a small time offset, we are confident that the ST5 discrepancies are due primarily to FAC or Pc2–Pc3 variability.

DMSP-ST5 discrepancies are smallest in the postterminator Southern Hemisphere polar cap, where disturbances from field-aligned currents tend to be minimal. Discrepancies tend to be larger in the Southern Hemisphere auroral zone where the conjunctions occur with greater altitudinal separations in regions of dynamic FACs and auroral arcs. It is possible that some Southern Hemisphere ST5-DMSP conjunctions show larger discrepancies because ST5 was in or above the field-aligned acceleration region. Our discrepancy statistics reveal a slight bias: DMSP perturbation magnitudes tend to ~20% larger than ST5 perturbation magnitudes (e.g., when ST5 measure a ~10 nT perturbation DMSP tends to measure a ~12 nT perturbation). This bias is under investigation.

We also investigated the influence that dayside ultralow frequency (ULF) waves, reported by *Le et al.* [2011] on closed field lines in the vicinity of noon, could have on our comparisons. Lu et al. showed Pc2–Pc3 waves with amplitudes of ~ 20 nT near 75 magnetic latitude. Such amplitudes, mapped in Modified Apex coordinates from ST5 altitudes to 110 km, will increase by ~10%, resulting in amplitudes of ~ 22 nT. Perturbations of this size may contribute to some Northern Hemisphere discrepancy between ST5 and DMSP; however, the ULF-caused discrepancy is likely small relative to other discrepancy sources.

Figure 1e shows that the magnitude of the interconstellation and intraconstellation conjunction discrepancies varies with geomagnetic activity level. The median values of discrepancy magnitude nearly doubles in the transition from quiet times ($Kp < 2$) to more active times ($2 \leq Kp < 4$). Both the interconstellation comparison (DMSP-ST5) and the intraconstellation comparisons (ST5-ST5) show similar trends. Interestingly, with increasing geomagnetic activity, magnetic perturbation vectors measured by the different spacecraft tend to have better directional agreement (not shown) likely because of the influence of stronger and more stable FACs.

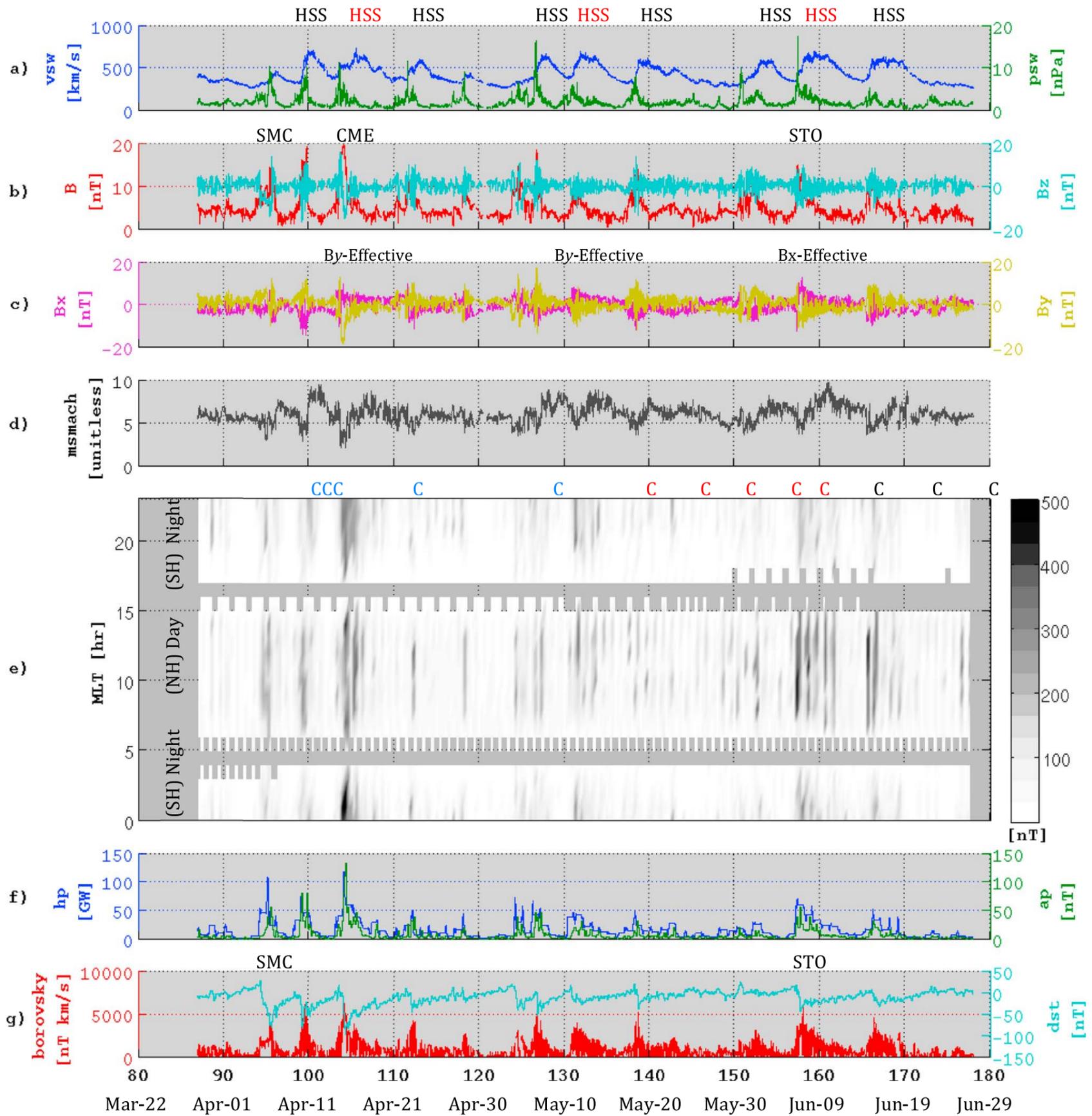

**Figure 2.** Overview of solar wind, magnetic perturbations, and geomagnetic index date for the ST5 mission, 25 March to 26 June 2006. The layout is (a) solar wind speed (blue) and dynamic pressure (green), (b) total IMF magnitude (red) and IMF $B_z$ (aqua), (c) IMF $B_x$ (magenta) and IMF $B_y$ (yellow), (d) solar wind magnetosonic mach number, (e) ST5 magnetic perturbations poleward of $+/-50°$ latitude bin-averaged over 8 h and 1 h magnetic local time (MLT). Dayside values and shading correspond to the Northern Hemisphere, and nightside shading and values correspond to the Southern Hemisphere; (f) hemispheric power (blue) and $a_p$ (green), (g) Dst (cyan) and the Borovsky coupling function (red). High-speed streams (HSS), steady magnetospheric convection (SMC), sawtooth oscillations (STO), and the five quietest days of each month (C) are marked.

The ST5-ST5 median discrepancy was ~30 nT while the median discrepancy magnitude between ST5 and DMSP was ~60 nT. The larger ST5-DMSP discrepancy likely has at two sources: (1) The "conjugate" position and time of measurement is neither truly cotemporal nor cospatial, given the 3° and ±60 s window; and (2) the perturbations are derived from an IGRF + baseline removal method that also has uncertainty. We note that *Alken et al.* [2014] who had access to DMSP data with ~30 m location accuracy along with high-accuracy

Oersted data, showed ~20 nT date-model discrepancies in the high-latitude regions. Given that we have access to ST5 and DMSP data with at best 1 km accuracy, we assert that the ST5 and DMSP measurements are in very good agreement.

## 3. Data Representation and Geophysical Context for the Full-Mission Magnetic Data

To show the ST5 reprocessed magnetic data in a geomagnetic context, we provide a magnetic keogram, shown in middle of Figure 2. Figure 2 gives an overview of solar wind data, the ST5 magnetic perturbation data, and geomagnetic indices, and derived parameters for the mission interval. The NASA OMNIWeb solar wind speed and density data (Figure 2a) show a repeating triplet of high-speed streams (HSSs) that dominated the interval (see *Thayer et al.* [2008] for a synopsis of the 2006 high-speed streams). Accompanying the HSSs are prestream interplanetary magnetic field (IMF) enhancements from corotating interaction regions (CIRs), and intrastream IMF fluctuations (Figure 2b), as well as IMF sector variations in the $B_y$ and $B_x$ components (Figure 2c). Figure 2d shows the solar wind magnetosonic Mach ($M_{MS}$) number—low values of this parameter indicates strong magnetic field in the magnetosheath and the potential for high reconnection efficiency [*Lavraud and Borovsky*, 2008; *Grocott et al.*, 2009]. A modest, but notable, decrease in $M_{MS}$ to values < 5 accompanies all of the CIRs and the coronal mass ejection (CME) on 13–14 April 2006. The lowest $M_{MS}$ value ($M_{MS} = 3.5$) of the interval was recorded during the CME passage on 15 April 2006.

In Figure 2e the ST5 magnetic perturbations poleward of ±50° latitude are shown as a keogram, with data bin-averaged over 8 h and 1 h magnetic local time (MLT), and plotted in a magnetic local time versus day of year format. Usually all three ST5 satellites contribute to the average; however, in the latter part of the mission there are several instances when only two satellites were available. With all data mapped to 110 km we are able to directly compare the intensity of dayside and nightside FAC perturbations. Magnetic perturbations from ST5 dayside passes (05–20 LT), with significant high-latitude coverage, are shown in the center section of Figure 2e. The dark stripes and patches show when FACs were enhanced. Note that ST5 northern passes provided near-cusp coverage for about half of each day. For the other half of the day ST5 passes were equatorward of the typical cusp locations. The result is a diurnal effect (more evident during high magnetic activity in Figure 2e) with alternating dark streaks indicating averaged high-latitude FAC perturbations, and light streaks indicating minimal FAC perturbations. Nightside passes (17–04 LT) that cover much of the southern auroral zone, except for the premidnight region, bracket the dayside passes. Dark gray regions without data show were the ST5 satellites transit middle and low latitudes.

Figure 2f displays 8 h boxcar averaged NOAA Polar Operational Environmental Satellite (POES) auroral hemispheric power [*Fuller Rowell and Evans*, 1987] overlaid on the 3 h $a_p$ index. Figure 2g contains the hourly *Dst* and the 5 min cadence Borovsky reconnection coupling function [*Borovsky*, 2013]. All of these generally match the magnetic disturbance levels of Figure 2e.

Labeling in Figure 2 highlights HSSs, the lone CME, extended intervals of magnetic calm indicated by "C," a steady magnetospheric convection (SMC) event, and a sawtooth oscillation (STO) event. Three of the HSSs are labeled in red. These streams had "effective" IMF orientations either due to Russell-McPherron IMF $B_y$ effects [*Russell and McPherron*, 1973; *McPherron et al.*, 2009 and *McGranaghan et al.*, 2014] or due to solstice dipole-tilt $B_x$ effects that we discuss in the next section.

## 4. Results and Discussion

### 4.1. FAC Disturbances

The keogram view of FAC-related magnetic perturbations shows that geospace responded to numerous solar wind disturbances during the ST5 mission. In the first half of April 2006 there were three strong events [*Rao et al.*, 2012] each with a different type of leading disturbance. The first event on 5 April (Day 95) developed as the IMF turned steadily southward ($B_z \leq -5$ nT) for 15 h ahead of a modest speed stream [*Veenadhari et al.*, 2012]. Steady magnetospheric convection dominated for several hours ahead of this CIR (see SMC list in *DeJong* [2014]). On 9 April (Day 99) a sharp pressure pulse with high solar wind speeds accompanied the second event. The event on days 103–104 (13–14 April) was produced by a CME just ahead of a HSS [*Le et al.*, 2009]. The strongest nightside perturbations (>1000 nT) and largest DMSP-ST5 discrepancy (~750 nT) accompanied this event.

Subsequent disturbances during the mission were primarily the result of CIR-HSS systems [*Verbanac et al.*, 2011]. Two CIR/HSS events near solstice, beginning on 7 June (Day 158) and 16 June (Day 167), registered unusually strong dayside FAC responses. The magnetosphere began a sawtooth oscillation (STO) during the slow flow ahead of the CIR on 7 June and continued in this mode through the CIR and into the early hours of the HSS (see STO list in *Cai and Clauer* [2009] and *Brambles et al.* [2013] for a discussion of STO drivers). Immediately following the STO event was a series of strong irregular substorms that formed an interval of high-intensity long-duration continuous *AE* activity [*Tsurutani and Gonzalez*, 1987]. The high-resolution *AE* index (not shown) clearly records the magnetotail response to the solar wind driving. ST5, on the other hand, has an unusually well-located and consistent view of the northern dayside high-latitude region during the events. The unusually strong dayside response, in relation to other CIRs in the interval, is notable in Figure 2e. The multiday response is likely the result of several factors. It is possible that the large dipole tilt during this interval allowed the positive IMF $B_x$ component to project onto the near-cusp northern magnetosphere fields in a manner that promoted magnetic merging closer to, but still equatorward, of the cusp [*Alexeev et al.*, 1998; *Maynard et al.*, 2003; *Tang et al.*, 2013]. Further, the accompanying IMF $B_y$ fluctuations likely extended the near-cusp FAC in longitude [e.g., *Li et al.*, 2011], thus allowing ST5 an unusually consistent view of the northern dayside high-latitude FAC region during a period of seemingly moderate HSS forcing. All of these interactions occurred when solar-driven conductivity was at its peak. The initial interval of the CIR/HSS on 16 June was subject to similar conditions; however, the IMF $B_x$ component became negative and less conducive to enhanced merging after the HSS flow developed.

### 4.2. FAC Calms

The white regions of Figure 2e, indicated by C, provide a unique, view of the near-Earth FAC system in its minimally perturbed state ($A_p \leq 7$). These include the five quietest days from each of the months of April–June 2006 determined from the *Kp* index (GeoForschungsZentrum (GFZ) Potsdam): 30, 12, 1, 3, and 2 April; 16, 21, 27, 29, and 9 May; and 26, 4, 23, 21, and 13 June. These dates are associated with relatively low solar wind speeds of ≤450 km/s and reduced levels of IMF magnitude and variation. Some of these events cover both dayside and nightside (events beginning on 30 March and 30 April), while others have minor levels of dayside activity with the nightside remaining quiet. Additional common elements of these events include Borovsky reconnection coupling values less than 500 nT km/s, and a northward IMF component, or a negative IMF $B_y$ component that projects positively onto Earth's dipole during the vernal equinox [*Russell and McPherron*, 1973]. During the magnetic calms the 8 h averaged FAC perturbations are < 100 nT. Intervals of general "calm-before-the-storm" have been noted by *Borovsky and Denton* [2013] and *McGranaghan et al.* [2014] as features that occur in the slow speed flow in the vicinity of the current sheet ahead of helmet streamer-driven HSSs. Both studies reported that these calms exert substantial control over the ensuing magnetospheric and thermospheric storm response.

### 5. Conclusions

The ST5 Technology Demonstration mission provided high-resolution magnetic data from three spacecraft in elliptical, low Earth orbits. The ~5000 polar passes span an interval of primarily low geomagnetic activity that was interrupted by a single strong geomagnetic storm early in the mission. We have mapped the magnetic perturbation data to 110 km via Apex transformations and provided them in common data format to NASA CDAWeb for community use. The mapping allows easy intercomparison of magnetic perturbations from both hemispheres. Gain jumps have been removed, and comparisons made to DMSP data in proximate locations. The DMSP and ST5 perturbations when compared at 110 km are in very good agreement, typically within 60 nT magnitude. This is well within the expected range given that the measurements are made at only approximately conjugate locations (within 3° and ±60 s).

We created a magnetic keogram to provide a comprehensive view of the magnetic response to different solar wind drivers over the life of the mission. This view shows expected responses to equinoctial high speeds streams and solar wind transients but also reveals magnetic calms associated with slow solar wind prior to high-speed streams and unexpectedly strong summer dayside high-latitude currents during the last weeks of the mission. We postulate that the latter response is associated with a solstice high-speed stream with a positive IMF $B_x$ component. Additional modeling studies will be needed to determine the relative roles of IMF $B_x$ and $B_z$ during such events and the locations of magnetic merging that support the enhanced cusp currents.

**Acknowledgments**

D.J.K., L.M.K., and R.J.R. were partially supported by NASA grant NNX13AG07G. DJ.K. was also partially supported by NSF grant AGS 1144154. L.M.K. was partially supported by AFOSR award 12–091; FA9550-12*0264. We benefited from conversations with Fred Rich of MIT, Dan Ober and Gordon Wilson of the U.S. Air Force Research Laboratory, and Arthur Richmond of NCAR. The ST5 data are available at http://cdaweb.sci.gsfc.nasa.gov/cdaweb/sp_phys/. The hemispheric power data are from http://www.swpc.noaa.gov/ftpdir/lists/hpi/power_2006.txt. The solar wind plasma data, IMF values, and the *AE* and *Dst* indices are available from NASA OMNIWeb. The Borovsky coupling function and magnetosonic Mach number were calculated from the NASA OMNIWeb data. Code for Apex magnetic conversions is at http://www.ngdc.noaa.gov/geomag/geom_util/apex.shtml. The DMSP conjunction data are in the supporting information to this manuscript. Higher resolution versions of Figure 2e and associated code (proprietary) can be requested from the first author. We are grateful to GeoForschungsZentrum (GFZ) Potsdam for computing the International Quiet Days from the *Kp* index. NCAR is sponsored by the National Science Foundation.